# Mechanical characterization of pharmaceutical powders and correlation with their behavior during grinding

L. Baraldi[1], D. De Angelis[1], R. Bosi[1], R. Pennini[1], I. Bassanetti[1], A. Benassi[1,2] and G.E. Bellazzi[1,*]

1 - Chiesi Farmaceutici SpA, Parma (Italy).

2 - International School for Advanced Studies (SISSA), Trieste (Italy).

*Corresponding Author address: Chiesi Farmaceutici S.p.A. Largo Belloli 11A– 43122 Parma (Italy)
email address: g.bellazzi@chiesi.com  phone: +39 0521 1689463

## Abstract

Controlling the size of powder particles is pivotal in the design of many pharmaceutical forms and the related manufacturing processes and plants. One of the most common techniques for particle size reduction in process industry is powder milling, whose efficiency relates to the mechanical properties of powder particles themselves. In this work, we first characterize the elastic and plastic response of different pharmaceutical powders by measuring their Young modulus, the hardness and the brittleness index via nano-indentation. Subsequently, we analyze the behavior of those powder samples during comminution via jet-mill at different process conditions. Finally, the correlation between single particle mechanical properties and milling process results is illustrated; the possibility to build a predictive model for powder grindability, based on nano-indentation data, is critically discussed.

## Introduction

Characterizing single particle mechanical properties can be of paramount importance in the design of many manufacturing operations involving pharmaceutical powders or in the excipient selection during early formulation of new drug products. Just to name a few examples:

- To have a good chance to reach the deep acinar airways, the size of the active pharmaceutical ingredient (API) particles in a dry powder for inhalation must lie between 1 and 5 $\mu m$ [1,2]. This range of particle size is typically attained by milling a coarser, brittle, starting material in a jet-mill, where particle collisions and consequent fragmentation are responsible for the size reduction [3–5]. Size reduction of API particles by milling is also frequently used to increase the dissolution, and thus the bio-availability, of poorly soluble drugs by enhancing the specific surface area accessible to the solvent [6,7];
- the ductility of API and excipient particles is very important during powder compaction, it determines the mechanical properties of the compact and thus the overall quality of oral solid product (tablet) [8,9];
- finally, a poor mechanical resistance of composite particles, such as soft pellets or granulated particles, might lead to alterations in the drug product quality during conveying and handling operations [10].

Establishing a correlation between single particle mechanical properties and the behavior of powder products during manufacturing operations might promote novel effective design methods. Cheap and fast measurements on single particle mechanical properties could be used to exploit new targeted materials and envisage innovative product solutions, thus reducing significantly time and material consuming trials in the plants.

Process design and control are nowadays increasingly assisted by numerical modelling and simulation[11–13]. For the realistic modeling of those processes where particle breakage occurs, both as a desired [14–16] or detrimental effect [17], a breakage probability distribution for the mother particle and a fragmentation distribution describing the generation of daughter fragments must be properly defined [18]. Meier et al. demonstrated how the parameters appearing in such functions directly depend on the single particle mechanical properties [19,20].

The nano-indentation technique allows to measure single particle mechanical properties with practically no consumption of material, e.g. only few tens of particles are necessary. Such versatility is particularly appreciated when dealing with API of new synthesis whose total amount, in the early development phases, is limited while the synthesis costs are considerable.

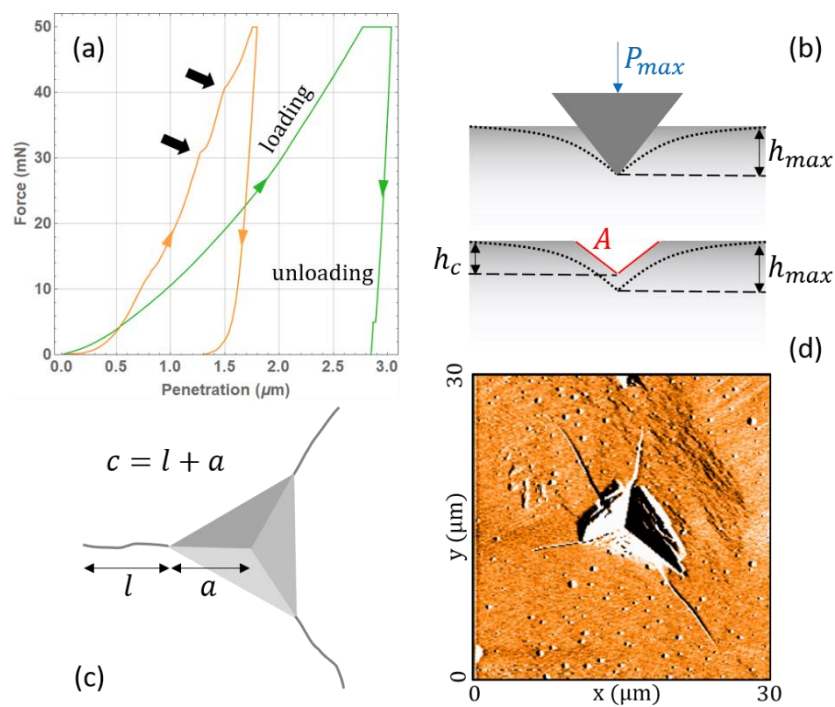

Figure 1: (a) mechanical hysteresis curves for sodium chloride (green) and lactose (yellow), black arrows indicate step-like features in the ascendant part of the loop due to crack propagation. (b) sketch of tip penetration during indentation and footprint depth. (c) sketch of the top view of a footprint with crack length measure. (d) AFM image of an indentation footprint on a compound A particle.

A typical nano-indentation test consists in pushing a hard metallic tip on a single powder particle, glued on a sample holder, with increasing force, while measuring the tip displacement [21–23]. Typical indentation profiles acquired on our samples are shown in figure 1 (a). Part of the tip displacement is due to the elastic deformation of the particle surface and part to plastic penetration inside it, as illustrated in figure 1 (b). From the recorded curve it is thus possible to extract information on both the elastic response and plastic (irreversible) deformation. By imaging the indentation footprint with optical microscopy, atomic force microscopy (AFM) or scanning electron microscopy (SEM), it is possible to measure the length of the generated cracks, an example is shown in panels (c) and (d) of figure 1. Assuming a certain model of deformation and crack propagation, beneath the tip, a brittleness index can be defined indicating the nature of plastic deformation, i.e. whether the material responds in a brittle or ductile fashion.

Naturally, single particle nano-indentation has also some intrinsic limitations. Powder particles must be tightly glued to a sample holder to be manipulated and placed individually under the indenter. This is only possible as long as the particles have a size of at least of 50 - 100 $\mu m$, below such limit cohesion forces will dominate over gravity, keeping the powder particles aggregated. Powders with a peculiar crystal habit (e.g. acicular, lamellar) will lie on the sample holder with certain preferential orientations, i.e. certain crystallographic directions might not be accessible to indentation. In other cases the single crystal shape (e.g. octahedra) might prevent the glued particles from exposing horizontal surfaces, i.e. planes parallel to the sample holder and perpendicular to the indentation axis, if this alignment condition is not met, indentation data are meaningless. If the indentation footprint is too small a local or "microscopic" value of the elasto-plastic properties of the material is probed, thus a large variability in the measures must be expected due to surface inhomogeneities, local disorder and defects. Only if the indentation footprint is large enough to average over the microscopic surface disorder and to probe enough "weak spots" of the particle surface, the data variability will be substantially reduced and the results will no longer depend on the indentation depth (meaning also maximum indentation force and footprint size) [23].

Several works are present in literature discussing the application of single particle nano-indentation to active pharmaceutical ingredients and excipients [24–26] and its correlation with powder processability [19,21,27–29]. Shariare et al. [25] demonstrated the difficulty in correlating directly the brittleness properties of pharmaceutical powders with the calculated interaction energy between crystalline planes within the particles. This difficulty stems from the fact that material breakage is dominated by defects, dislocations and pre-existing flaws which might lower considerably the theoretical energy barrier to initiate slip motion between molecular planes, as proved by Vegt et al. [29,30]. Zuegner et al. [28] found that the connection between powder milling efficiency and the elastic and plastic properties of single particles is non-trivial. They concluded that particle elasticity impacts the breakage resistance of materials more than particle hardness. However their picture cannot explain the behavior of materials such as sodium chloride and, without any further measurement, it is not possible to discriminate whether the plastic deformation of particles has a ductile or brittle character. This problem was further addressed by Meier et al. [19] showing how the brittleness index remains the most promising parameter to predict the milling behavior of powders. Taylor at al. [27] found a very nice correlation between the milling efficiency and different single particle mechanical properties such as Young modulus, hardness and brittleness index of different pharmaceutical compounds. Single particle nano-indentation data correlates very well also with the powder behavior during compaction and tableting processes; a summary can be found in the review by Egart et al. [21].

In this work we characterize the elastic and plastic response of different pharmaceutical powders by measuring Young modulus and hardness via nano-indentation. The estimation of the average length of the generated cracks allows the evaluation of fracture toughness and brittleness index. Our results are compared with those already present in the literature, the variability and reproducibility of the measurements is also analyzed. Particle comminution is then studied through a tabletop jet mill, size reduction effects are discussed for the different materials and process conditions. Finally, the correlation between single particle mechanical properties and the powder grindability is illustrated and discussed, as well as the possibility to build a predictive model for the powder behavior during milling, solely based on single particle mechanical properties measured via nano-indentation.

## Materials and Methods

### Materials selection

APIs and excipient for the nano-indentation analysis have been selected based on three different criteria. First, our choice is limited by the sample abundance as we need few tens of grams of powder to perform jet-milling trials. Such quantities are not easily granted for new compounds whose synthesis reaction is still in its early phase design. A second important aspect is that, to validate our indentation method, we need some reference compound whose mechanical properties are already known and whose crystalline nature is similar to those materials we synthesize in-house. Commercial excipients have been selected for this purpose: sodium chloride and L-(+)-tartaric acid from Merck® and lactose $\alpha$-monohydrate from Armor-Pharma®. Sodium chloride and lactose have been both analyzed as they appear out-of-the-envelope and re-crystallized by slow evaporation from a saturated water (sodium chloride) or ethanol (lactose) solution. Finally, our material selection was also driven by the availability of suitable size and shape of the crystalline particles, large enough to be manipulated and glued exposing large and flat surfaces. Different particle/crystal morphologies have been explored, from irregular prisms, passing through large plates, to elongated homogeneous needles. While crystals of compound A have been directly used as obtained by bulk synthesis, for compounds B and C a re-crystallization procedure has been necessary to achieve larger particle sizes. For the single crystals of compounds A and B, the largest particles have been selected and inspected at the stereo microscope while for the bulk crystals of compound C a gentle sieving step with a 125 µm mesh has been necessary, before particle selection, in order to separate a pre-existing fine particle fraction.

### Particle size characterization

The particle size distribution (PSD) of the starting and micronized material has been determined using a Sympatec HELOS/BF instrument equipped with RODOS/ASPIROS dry dispersion unit. The powder dispersion has been performed at 2 bar pressure drop with a sled speed of 30mm/s. Different combinations of R1 (0.1/0.18 µm - 35 µm), R3 (0.5/0.9 µm - 175 µm) and R5 (0.5/4.5 – 875 µm) lenses allowed to cover the entire particle distribution range for the samples. The PSD characterization of the micronized material has always been performed within few hours from the milling, all samples have been stored in dry nitrogen atmosphere to avoid changes in their surface characteristics possibly leading to particle sintering. The presented PSD and dv90 data have been averaged over three replica of the analysis, with an estimated uncertainty below 1 $\mu m$.

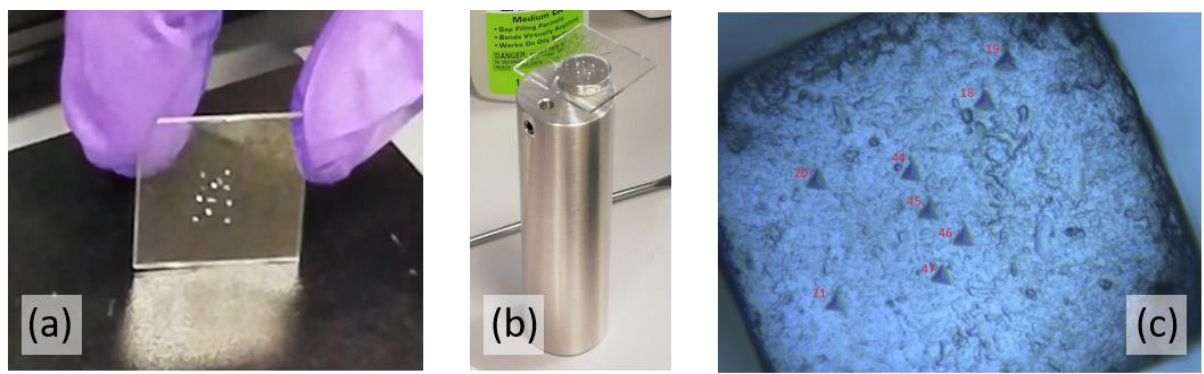

*Figure 2: (a) an example of glass slide with selected particles glued on it. (b) the glass slide mounted on the sample holder apparatus before the insertion in the indentation equipment. (c) some indentation footprints imaged by optical microscopy on a sodium chloride particle.*

## Indentation test

### Sample preparation

For each measurement session, particles were carefully selected using stereo and optical microscopy in order to find and isolate those sufficiently large, with smooth and uniform surfaces. Precision tweezers have been employed to glue the selected particles on a dedicated square glass slide where a thin film ($< 500\ \mu m$) of biphasic epoxy glue has been previously deposited. Particles are deposited in the central part of the glass slider only, as shown in figure 2 (a). After 15 minutes the glass slide has been very gently brushed to remove powders residues, dust agglomerates or particles not perfectly glued. The glass slide is in turn glued on the nano-indenter sample holder and left drying for at least 30 minutes, see figure 2 (b). Notice how the glued particles lie all within the area of metallic cylinder of the sample holder: indenting too far from it would add a spurious elastic response due to the glass slide bending. In figure 2 (c) an optical microscopy image shows several indentation footprints on the same sodium chloride particle.

### Indentation method

The indentation experiments have been performed using a NanoTest™ Vantage nano-indenter (from Micro Materials®) equipped with a sharp pyramidal Berkovich diamond indenter. The typical indentation protocol is illustrated in figure 3 (a), it consists of a loading ramp of $20\ s$ duration, a $30\ s$ dwell period at maximum load $P_{max}$ and a final unloading step lasting $10\ s$. The corresponding sample deformation profile is shown in figure 3 (b) and (c). Notice how the tip keeps penetrating the sample even when the loading force is kept constant, this occurs due to creep deformation. The purpose of the dwell period at maximum loading is precisely to let the creep rate reduce so that plastic deformation is minimized, or completely absent, during the subsequent unloading step, in which the elastic character of the sample is probed. At the very end of the unloading curve, with only 10% of $P_{max}$ still applied, a second dwell period of $60\ s$ is maintained to estimate the tip thermal drift and correct for this effect in the Young modulus $E$ and hardness $H$ calculation. For different materials and samples, the maximum applied load ranged from 5 to 200 $mN$ depending on the ease in the crack production without the generation of major damages to the particle, i.e. chipping or particle breakage.

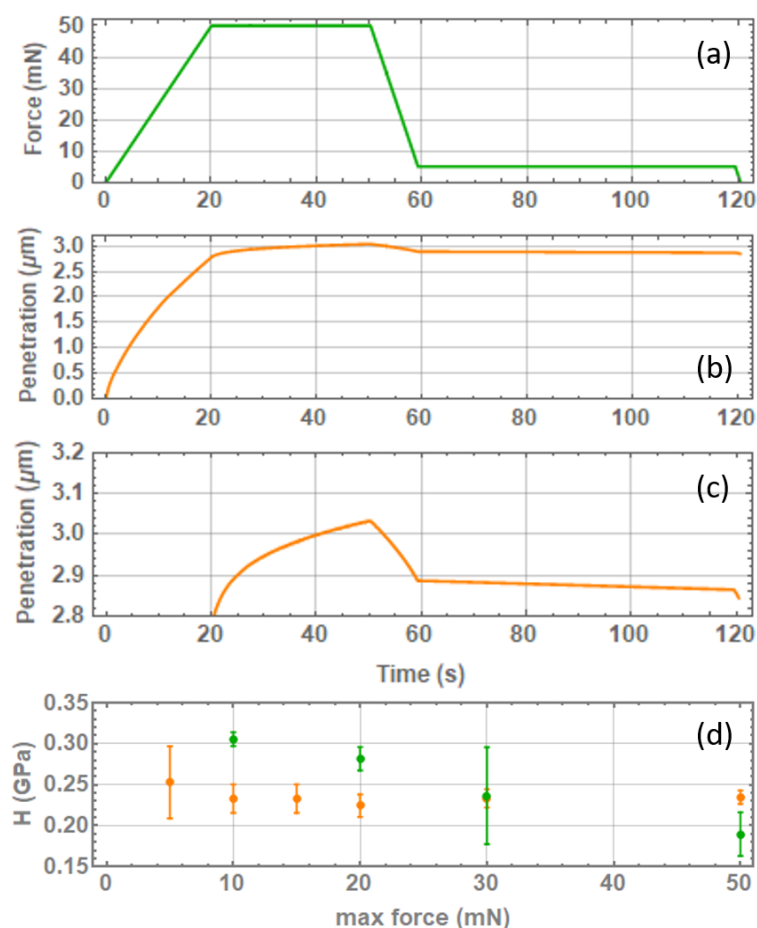

Figure 3: (a) Applied force as a function of time during a typical indentation cycle. (b) and (c) penetration depth as a function of time when the force profile of panel (a) is applied on a sodium chloride particle. (d) some example of how the measured mechanical properties depend on the maximum loading force $P_{max}$, for re-crystalized sodium chloride (yellow) and compound c (green).

## Constants determination

The indentation hardness $H$ is a measure of the attitude of a material to be irreversibly deformed by an external mechanical stress, exerted through a tip of known shape. With reference to figure 1 (b), $H$ is defined as:

$$H = \frac{P_{max}}{A(h_C)} \qquad (1)$$

where $P_{max}$ is the maximum force applied by the indenter tip on the sample surface and $A$ is the area of the generated indentation footprint. The latter is a function of the tip shape and penetration depth $h_C$, for the Berkovich tip used in this work the correlation between the footprint area and the penetration depth is given by $A = 24.56\ h_C^2$. Notice how, as an irreversible deformation of the sample surface occurs, the loading and unloading curve do not coincide, i.e. a mechanical hysteresis loop is recorded. The reason for using $h_C$, in the hardness calculation rather than $h_{max}$, is that the latter contains both the plastic and elastic deformation, as sketched in figure 1 (b).

From the slope of the unloading curve it is also possible to infer the Young modulus value $E$, representing how prone is a material to elastic deformation under the action of an external force:

$$E = \frac{\sqrt{\pi}}{2\beta} \frac{dP}{dh} \frac{1}{\sqrt{A(h_C)}} \qquad (2)$$

$\beta$ is a tip geometry constant ($\beta = 1.034$ for our Berkovich indenter [21]). As the recorded curves usually present some noise, the $dP/dh$ derivative cannot be easily computed numerically via finite-difference algorithms, rather the instrument software usually fit the unloading curve with some model expression of known analytical derivative [31]. Under the same external force, a material with small $E$ value will show a significant reversible deformation, e.g. rubber, a material with large $E$ will give rise to an infinitesimal elastic deformation, e.g. steel. Notice that $E$ appearing in eq.(2) is the reduced Young modulus:

$$\frac{1}{E} = \sum_i \frac{(1 - \nu_i^2)}{E_i} \qquad (3)$$

where $E_i$ and $\nu_i$ are the Young modulus and Poisson ratio of each deformable medium (in series) in the measurement system, i.e. the tip, the powder particle, the glue layer and the sample holder. $E$ coincides with the particle young modulus $E_p$ only assuming that the tip, the glue layer and the sample holder are orders of magnitude stiffer that the particle, thus remaining practically undeformed during the whole indentation process. Only in this specific situation:

$$\frac{1}{E} \cong \frac{(1 - \nu_p^2)}{E_p} \qquad (4)$$

As $\nu_p$ is very difficult to measure on single particles, the value of $E$ is usually reported in literature rather than $E_p$ [23], we have adopted the same convention throughout the rest of the paper. In any case, typical values of $\nu_p$ for crystals of organic molecules are $0.2 - 0.25$, for sodium chloride $\nu_p = 0.18$ [29] which means $E_p \sim 0.97\ E$, i.e. the difference between Young modulus and reduced modulus is smaller than the typical uncertainty in the measure itself.

At least two independent measurement campaigns have been performed for each selected material, employing different operators and with independent sample preparation to evaluate the method reproducibility and to test its sensitivity to sample preparation. For each $P_{max}$ value, 5 to 10 indentation tests have been performed over 2 or more particles. For some of the investigated materials we noticed the typical overestimation of $E$ and $H$, as well as their larger variability, when small $P_{max}$ values are applied, an example is shown for sodium chloride in figure 3 (d). For other materials we have noticed that both $E$ and $H$ values keep decreasing slightly with increasing $P_{max}$ and that the variability of the measure is not necessarily correlated to $P_{max}$, see the compound C curve in the same figure. Increasing further $P_{max}$ results in particle chipping or breakage thus preventing us from reaching a condition where $E$ and $H$ are completely independent from $P_{max}$. For this reason we decided to present our measurements as an average over all the applied $P_{max}$ range.

## Footprint Analysis

### Imaging Methods

By inspecting the indentation footprint it is possible to assess the nature of the irreversible deformation, i.e. whether the mechanical energy lost during the irreversible deformation is dissipated in a ductile displacement of material around the footprint or through brittle fractures. The fracture toughness $K_C$ is a critical stress above which the cracks, naturally present in a particle, start to propagate irreversibly thus breaking it into smaller fragments. Such value is calculated combining the previously defined mechanical properties $H$ and $E$ with the crack length $l$, or the other characteristic length $c$ defined in figure 1 (c), measured through optical or atomic force microscopy. Crack footprints have been investigated by optical microscopy (Nikon® Eclipse LV 100POL) and by atomic force microscopy (C3000 by Nanosurf®). The AFM probe has been operated using the imaging mode and in dynamic force acquisition/tapping mode with long cantilever 190AI-G. The Z-controller setpoint, the P/I gain, the tip voltage and free vibration amplitude were modified during the analysis and depending on sample nature in order to produce the most resolved image possible.

### Constants determination

Several models have been proposed for $K_C$ based on different assumptions on the deformation and crack propagation beneath the tip [32]. Meier et al. tested 19 of them on sucrose finding up to one order of magnitude variability in the results [19]. In this work we limit our analysis to the most commonly employed model of Laugier:

$$K_C = \gamma \sqrt{\frac{E}{H}} \frac{P_{max}}{c^{3/2}} \quad (5)$$

where $\gamma = 0.016$. Now a brittleness index $b$ can be defined as:

$$b = \frac{H}{K_C} \quad (6)$$

such ratio is very small either when the material is hardly plastically deformable (small $H$) or when the stress necessary to start crack propagation is very large ($K_C$ large), in both cases the material will not deform at all or it will undergo a ductile deformation. On the contrary, a large $b$ value is associated to small stresses to initiate crack propagation (small $K_C$) and thus to particle breakage by brittle fracture.

Another interesting quantity to estimate, once $K_C$ is known, is the critical diameter $d_c$, i.e. the characteristic particle size below which brittle fracture no longer occurs [33–35]. If fragmentation by brittle fracture would be the only size reduction mechanism, $d_c$ could be a measure of the smallest particle size attainable with any milling process. In fact, particles with diameter smaller than $d_c$, could only deform in a ductile manner under compressing loads. Kendall and Hagen proposed independently two different models for this ductile-brittle transition, obtaining for $d_c$ the same dependence on $K_C/H$:

$$d_c = \alpha \left(\frac{K_C}{H}\right)^2 \quad (7)$$

in this work we adopt the Hagen model where $\alpha = 29.5$, more details about the different model assumptions can be found in the work of Knieke [34].

Crack imaging and analysis is a time-consuming operation, we thus performed it only on a limited subset of the indentation sessions, at least one for each selected material. If not otherwise stated, the presented results are the average over at least 5 different footprints.

## Micronization

Size reduction has been performed through a LaboMill jet mill, from Food Pharma Systems®, equipped with a 1.5 inches PTFE milling chamber and 4 inert stainless-steel nozzles. The mill has been fed with 1.6 g of powder manually dosed regularly in time to maintain a constant feed rate of 10 g/h. Nitrogen has been selected as the milling gas. All the materials have been tested at different grinding pressures, namely 3, 4.5 and 6 barg. The feeding/injection pressure was set 1 barg above the grinding one to avoid blow-back phenomena.

# Results and discussion

The data on the elastic and plastic behavior acquired for all the samples are summarized in Table 1, the values present in literature for the reference compounds are also reported for comparison (sodium chloride [24,28], tartaric acid [19], lactose [19,28]). The hardness of Sodium chloride is found in good agreement with the data reported by Duncan-Hewitt et al. [24] while some discrepancy is found for the Young modulus. This material is known to be extremely ductile with a very limited elastic deformation, which makes the unloading curve of the hysteresis loop very steep and thus extremely hard to be correctly fitted to extrapolate its derivative by the instrument software. Testing different fitting formula and different numerical methods to determine the fitting coefficients, through an external software, we found $E$ can increase even by $5-7 GPa$. This brings the re-crystallized sample value in agreement with the experimental one and enhance slightly the commercial sample value which remains lower. Although with a little more variability between the different sessions, also the hardness of Tartaric Acid is close to the value reported by other authors; the Young modulus is found in good agreement except for the commercial session 2 where again we recorded an underestimation. Drops in the $E$ value could be justified by particle gluing issues, if the glue layer is not perfectly stiff and it undergoes a significant deformation, two terms of eq.(3) must be retained leading to $E > E_p$, in particular if the deformation of the glue layer is comparable to the particle one we get $E \sim E_p/2$. Both $E$ and $H$ of lactose are in reasonable agreement with the data reported in the literature, a slightly higher variability is visible for $H$ among the different sessions.

Table 1: Measured Young modulus and Hardness ± standard deviation. Comparison among different batches, different measurement sessions

| Material | Session | E (GPa) | H (Gpa) |
|---|---|---|---|
| Sodium Chloride | literature | 43.0-46.5 ± 1 | 0.213-0.44 ± 0.05 |
| | recristalized | 35.98 ± 3.0 | 0.23 ± 0.02 |
| | commercial | 22.83 ± 4.3 | 0.20 ± 0.03 |
| Tartaric Acid | literature | 43.34 ± 0.7 | 1.35 ± 0.05 |
| | commercial 1 | 37.96 ± 7.2 | 1.09 ± 0.21 |
| | commercial 2 | 32.18 ± 7.8 | 0.78 ± 0.18 |
| | commercial 3 | 42.11 ± 7.3 | 0.85 ± 0.33 |
| Lactose | literature | 21.44-23.7 ± 1 | 0.869-1.1 ± 0.06 |
| | recristalized | 27.59 ± 5.2 | 1.04 ± 0.15 |
| | commercial 1 | 23.38 ± 5.4 | 0.79 ± 0.3 |
| | commercial 2 | 21.44 ± 3.1 | 0.60 ± 0.18 |
| compound A | 1 | 9.04 ± 0.9 | 0.24 ± 0.03 |
| | 2 | 8.71 ± 1.3 | 0.28 ± 0.04 |
| | 3 | 12.41 ± 2.1 | 0.42 ± 0.06 |
| | 4 | 12.41 ± 1.1 | 0.42 ± 0.06 |
| | 5 | 8.19 ± 3.1 | 0.31 ± 0.15 |
| compound B | 1 | 8.19 ± 0.6 | 0.38 ± 0.04 |
| | 2 | 6.83 ± 0.8 | 0.30 ± 0.03 |
| | 3 | 4.76 ± 0.5 | 0.30 ± 0.02 |
| compound C | 1 | 6.64 ± 0.4 | 0.28 ± 0.03 |
| | 2 | 7.21 ± 2.4 | 0.38 ± 0.1 |

For sodium chloride and lactose it is possible to assess the effect of re-crystallization on the measured mechanical properties. For both materials Young modulus and hardness are found to be larger than for the out-of-the-envelope counterpart. We are thus led to the reasonable conclusion that a small-scale, controlled re-crystallization makes the powder particles harder and more elastic. This finding is in agreement with the results published by de Vegt et al. [29], showing how an increasing density of pre-existing flaws lowers both $E$ and $H$. The contained variability among different measurement sessions for compound B and C reveals a good reproducibility of the gluing procedure and a weak dependence on the operator. The same is true for sessions 3 to 5 of compound A, all performed on the same batch, while a pronounced difference is visible comparing with test 1 and 2 performed on two other different batches, although obtained from similar synthesis processes. This finding highlights the importance of characterizing the mechanical properties of every single manufactured batch before setting the milling process specifics and, in any case, warns about the intrinsic batch-to-batch variability that, for certain molecules of new synthesis, can be significant.

The crack length $c$ has been determined only on a subset of the measurements presented in Table 1, $K_C$, $b$ and $d_C$ have been subsequently calculated, results are summarized in Table 2. Sodium Chloride particles are extremely ductile and, even applying a high maximum loading $P_{max} = 500\ mN$, we never succeeded in inducing cracks around the indentation footprints. This behavior has been discussed also by other authors with the exception of Duncan-Hewitt et al. [24] obtaining the values reported in Table 2. Indeed their $K_C$ value is very high compared to all the other materials here analyzed, confirming no attitude to brittle fracture, i.e. $b \to 0$. For tartaric acid our $K_C$ and $b$ values compare nicely with literature data while lactose results are in good agreement for what concerns $K_C$ and revealing a smaller value for $b$. As already noted for $E$ and $H$, compound A shows a considerable batch-to-batch variability, for session 3 only one indentation produced good cracks thus no standard deviation is reported. Also the three sessions on compound B give consistent and reproducible measures. Compounds B and C appear to be the most brittle, they are thus expected to be easily grindable in milling processes.

Table 2: Measured crack length, fracture toughness and brittleness index ± standard deviation. Comparison among different batches, different measurement sessions and literature data.

| Material | Session | c (µm) | Kc (Mpa m^1/2) | b (m^-1/2 *1000) | dc (µm) |
|---|---|---|---|---|---|
| Sodium Chloride | literature | / | 0.5 ± 0.07 | 0.426 | 70.35 |
|  | recristalized | 0 | / | 0 | / |
|  | commercial | 0 | / | 0 | / |
| Tartaric Acid | literature | / | 0.17 ± 0.01 | 8.11 | 0.46 |
|  | commercial 3 | 9.5 ± 2.1 | 0.18 ± 0.07 | 9.10 | 0.6 |
| Lactose | literature | / | 0.09 ± 0.03 | 9.55 | 0.26 |
|  | commercial 2 | 11.88 ± 1.48 | 0.14 ± 0.02 | 5.92 | 0.94 |
| compound A | 1 | 16.41 ± 6.32 | 0.11 ± 0.08 | 3.60 | 9.6 |
|  | 2 | 21.50 ± 5.88 | 0.05 ± 0.02 | 6.52 | 1.47 |
|  | 3 | 13.13 ± n.a. | 0.035 ± n.a. | 11.09 | 0.24 |
|  | 5 | 17.15 ± 3.27 | 0.027 ± 0.006 | 16.08 | 0.12 |
| compound B | 1 | 5.5 ± 0.61 | 0.030 ± 0.004 | 13.85 | 0.17 |
|  | 2 | 10.71 ± 2.96 | 0.024 ± 0.005 | 13.09 | 0.22 |
|  | 3 | 7.5 ± 2.02 | 0.03 ± 0.006 | 11.90 | 0.28 |
| compound C | 2 | 16.15 ± 2.93 | 0.025 ± 0.004 | 17.64 | 0.09 |

The standard deviation associated to all our measurements is higher than the one reported by other authors for commercial excipients. This is in part due to our choice of averaging over all the applied loads, including the small ones usually enhancing the measure variability. A deep investigation on how the indentation protocol might affect the results variability is certainly necessary, in particular the impact of loading and unloading rates as well as the importance of the dwell time must be assessed. This point is poorly discussed in the literature, it would require a large measurement campaign, focusing on few materials, and is beyond the scope of the present paper.

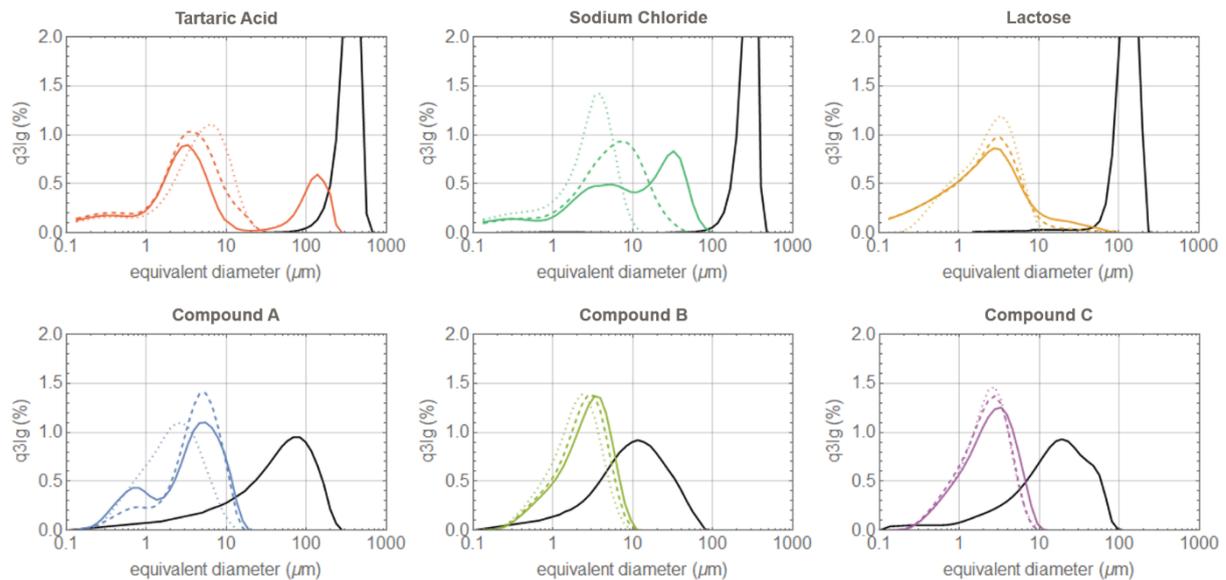

*Figure 4: PSD of the different materials before and after grinding. The black lines represent the starting material PSD, colored continuous, dashed and dotted lines represent the grinded product PSD at 3, 4.5 and 6 barg grinding pressure respectively.*

Both the synthesized compounds and the commercial excipients (in their out-of-the-envelope form) have been milled at the same powder feed rates and grinding pressures. PSD of the processed material has been acquired immediately after milling, the results are summarized in figure 4. Compound C is indeed the most easily grindable, at 3 barg its PSD is already close to the limit distribution with a dv90 well below 10 µm. Further increases in the grinding pressure only slightly sharpen it without any significant left shift. The same applies for compound B with only a minor shift observed with increasing pressure. Compound A is harder to mill, at low pressure collisions are not so effective in reducing the particle diameter, as a consequence the particle residence time is high, i.e. a large hold-up mass accumulates in the mill chamber reducing even further both the nitrogen and particle velocity [36]. At small collision velocities chipping and abrasion mechanisms are known to onset, generating small size fragments which could lead to the observed bi-modal size distribution [37–39]. Another possible scenario justifying a bi-modal distribution at low grinding pressure is an intermittent grinding regime [36]: if the product is difficult to grind, particles accumulate in the milling chamber, the large hold-up mass slows down the fluid and its centrifugal force, until most of the particles are able to reach the classifier and escape the mill although still large in size, giving rise to the rightmost peak. Now the mill is almost empty, the fluid accelerates promoting high energy collisions, very effective in size reduction, and preventing large particle to escape by restoring a stronger centrifugal force. In this way only small size particles escape the mill giving rise to the leftmost peak, this optimal milling condition is gradually lost as the milling chamber refills again. Only at 6 barg collisions are effective in provoking brittle fracture, this restores a mono-modal PSD slightly left-shifted. Unfortunately the milling equipment does not allow for further pressure increase to verify if such distribution is already the limit one. Something similar happens for lactose but here it is clear the limit distribution is reached already at 4.5 barg. Sodium chloride also displays a bi-modal distribution at low pressure which disappears with increasing pressure as the PSD shift to the left and sharpens. The mill is clearly working in an intermittent regime also for tartaric acid and, during its periodic emptying, particles exit with almost the same size they entered it. Increasing the grinding pressure the intermittent regime disappears but the main peak shifts to the right. This can be due to two phenomena: with higher collision speed, chipping and abrasion mechanisms are suppressed and brittle fracture produces only larger fragments; alternatively, a regime where milling efficiency decreases with increasing grinding pressure is known to occur at high feed rates [36], the milling pressure should be increased even more to see the PSD peak shifting back to the left. One might argue that the increasing difficulty in grinding certain materials (sodium chloride and tartaric acid in particular) is due to the larger particle size of the feed rather than to a less brittle attitude. To demonstrate that this is not a correct interpretation we sieved the tartaric acid material removing the coarser fraction from the feed. The difference between the original feed and the sieved one can be appreciated comparing panels (a) and (b) of figure 5. Probably due to the fracture of some weak large particles, the new feed shows a left tail enhancing the content of particles with diameter between 10 and 100 $\mu m$, the overall feed PSD resembles closely the lactose one. However the milled PSD and the powder behavior with increasing grinding pressure remains exactly the same of the coarser feed.

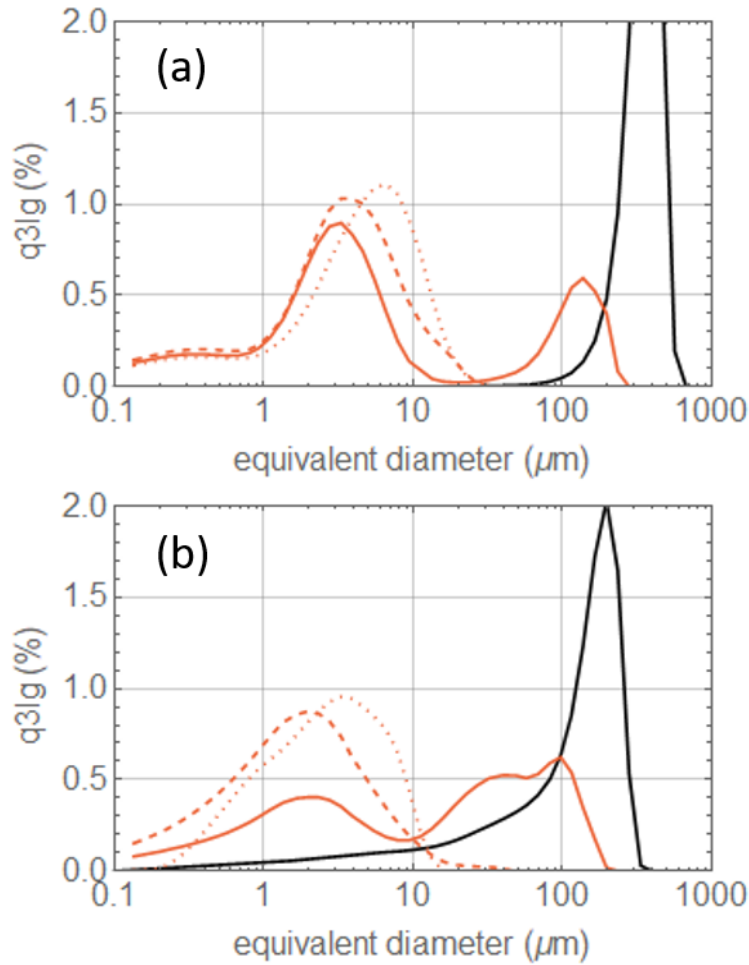

*Figure 5: PSD of tartaric acid for different grinding pressures with a coarser (a) or finer (b) starting material. The black lines represent the starting material PSD, colored continuous, dashed and dotted lines represent the grinded product PSD at 3, 4.5 and 6 barg grinding pressure respectively.*

Reducing the effect of the hold-up and accessing higher grinding pressures, to reach the limit PSD also for hard materials, could be both achieved using a larger mill. However, working with a larger grinding chamber implies the use of bigger powder samples, moving from few grams to tens of grams per trial, such amount of powder is generally not available in the early stages of new drug synthesis.

In setting the milling process parameter for a new material we are primarily interested in:

(i) understanding if the particle size can be reduced below a certain value;
(ii) how high must we rise the grinding pressure to achieve the desired PSD. As this depends also on the powder feed-rate we should talk about specific milling energy, however all the

- (iii) how strongly the PSD of the milled material depends on our choice of milling pressure;
- (iv) estimating the smallest particle size attainable, as this determines the span of our PSD.

presented results have been collected at the same feed rate thus the specific milling energy varies with pressure only;

To investigate point (i) we plotted the $dv90$ of the milled samples against their brittleness index for different grinding pressures, see figure 6 (a). Given the impossibility to generate cracks on sodium chloride, we used the $b$ value from Duncan-Hewitt et al. [24]; for compound B we averaged $b$ over the data from the different sessions; for compound A, where a large batch-to-batch variability is present, we used only the $b$ value measured directly on the milled batch. With the exception of tartaric acid, whose strongly bi-modal distribution makes the $dv90$ of little use, there is indeed a correlation. 3 barg data appear more scattered as the low-pressure PSD retains memory of the starting material one, at 6 barg most of the materials reach the limit distribution and the differences in $dv90$ gets thin. Fitting the data it is possible to estimate a dependence of $dv90$ on the measured brittleness index $b$, for the 3 and 6 barg data we get respectively:

$$dv90 = 5.8 \, e^{-b/51.6} \quad (8)$$

$$dv90 = 30.9/b^{0.43} - 4.0 \quad (9)$$

whose predictivity is of course valid only for a certain milling pressure and our specific mill geometry. More general expressions with an explicit dependence on pressure could be defined collecting data on many more materials and at different pressures, they could be used to answer to point (ii), i.e. to predict the $dv90$ of a new material just estimating its $b$ via nano-indentation. Looking at panel (a) of figure 6 it is also clear how the difference in $dv90$ between 3 and 6 barg becomes narrower as $b$ increases. Very brittle materials should in fact reach their limit distribution already at very low pressures while harder materials loose memory of their starting material PSD only at very high pressure. Thus, similarly to what other authors did, we propose the definition of an index representing the ease of milling a powder, a sort of grindability index, as $g = (dv90_{3barg} - dv90_{6barg})/ dv90_{3barg}$. Being estimated comparing the PSD of two milled samples, rather than referred to the starting material PSD, this index is by construction suitable to answer to point (iii). The grindability index $g$ as a function of $b$ is plotted in panel (b) of figure 6. Again with the exception of tartaric acid all the points lie on the fitting curve

$$g = 0.037\, b + 0.21 \quad (10)$$

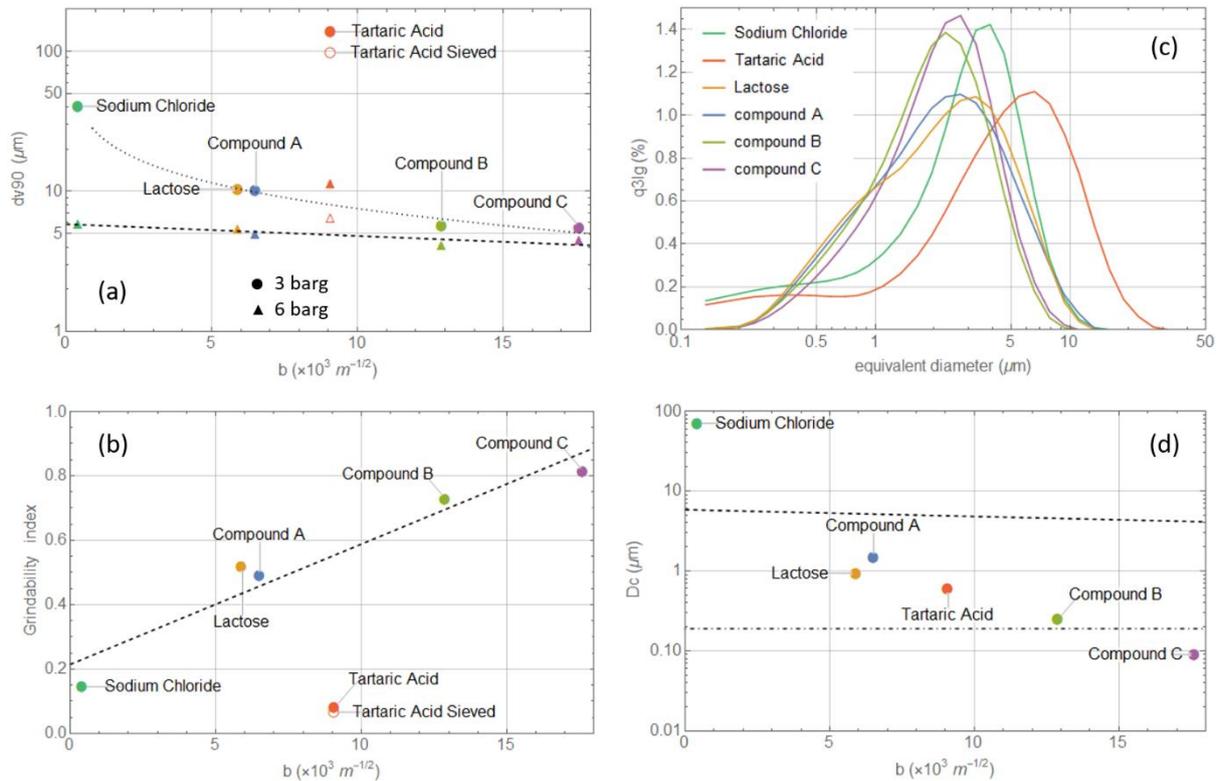

Figure 6: (a) dv90 as a function of the brittleness index b for 3 and 6 barg milling pressure (circles and triangles respectively), the black dashed and dotted lines represents the fit of eq.(8) and (9). (b) grindability index as a function of the brittleness index b, the black dashed line represents the fit with equation (10). (c) overlay of the PSD of the different samples milled at 6 barg grinding pressure. (d) critical particle diameter estimated through eq.(7) (colored dots), average diameter of the smallest fragments as measured from the PSD of panel (c) (dot-dashed line) and eq.(8) fit (dashed line, same as panel (a)).

Which again has a predictivity limited to a certain pressure range and bounded to our specific milling equipment. Why tartaric acid does not follow the general trends here highlighted for the other materials remains unclear. According to its $b$ value it should be a medium brittleness material, however, its behavior in the mill suggests an extreme difficulty in size reduction with accumulation of a large hold-up mass. Its large Young modulus, associated to large feed particles, might change completely the particle dynamics inside the mill: random frequent elastic rebound lower significantly the average particle-particle and particle-wall velocity reducing the comminution efficiency by brittle fracture. However this mechanisms should be also at work for sodium chloride, it also has similarly large $E$ and even less attitude to brittle fracture, and still it is found to be more easily grindable than tartaric acid.

Panel (c) of figure 6 shows the overlay of the PSD for all samples milled at 6 barg, assuming these distribution closely resemble the limiting ones (which is verified only on compound B and C and for lactose) one can verify the predictivity of the Hagen model and try to answer to point (iv). For compound A, B and C as well as for lactose the smallest fragment size is around 0.2 µm, for sodium chloride and tartaric acid longer left tails suggest the presence of much smaller fragments. This is however a numerical spurious effect emerging when combining the measurements of different lenses to generate a unique PSD. Checking the PSD of the sole R1 lens we found also for sodium chloride and tartaric acid a minimum fragment size of 0.2 µm. The smallest fragment size attainable by brittle fracture as predicted by the Hagen theory, is calculated through eq.(7), reported in Table 2 and plotted in figure 6 (d). Predicted and measured data are in good agreement only for lactose and compound B. The prediction for compound A, tartaric acid and sodium chloride significantly overestimate the smallest attainable particle size by orders of magnitude, an underestimation occurs for compound C. We must conclude that, for these compounds, either the simple assumptions underlying the Hagen model are not correct or brittle fracture is not the sole size reduction mechanism at play. The inadequacy of the Hagen model has been reported also by other authors showing how single particles can break into fragments much smaller than the calculated $d_c$ [34]. Their argument is that the simple models for the estimation of $d_c$ are based on static stress distributions in the samples, being thus more suitable for quasi-static loading experiments rather than for high speed collisions such as those occurring in single impaction testers (or jet-impact tester) as well as in jet-mills. Further investigations in this direction would require a dedicated work and the use of a particle size analysis tool more suitable for the sub-micron range, i.e. application of Mie light scattering theory with wet dispersion laser diffraction method.

## Conclusions

Once the proper sample preparation procedure has been set up, the indentation analysis proved to be reliable and reproducible. The measurements on commercial excipients are in good agreement with the data available in literature. However, the single particle handling and gluing, the search for good and flat terraces for indentation and the measure of the crack length are all manually executed operations which make of nano-indentation a very time-consuming analysis. Another main weakness of the technique is the need for re-crystallization, to generate larger particles for those compounds whose selected synthesis path leads to the precipitation of very small crystals. This is not only another time-consuming procedure, but it might alter significantly the mechanical properties of the particles, making them not representative of the larger amount of powder used in the milling trials. The differences in the mechanical properties of re-crystallized and out-of-the-envelope commercial excipients are evident from our analysis, however re-crystallization did not prevent us from obtaining a good correlation between $dv90$ and $b$ for compound B and C. This might mean that, for these compounds, the re-crystallization process does not alter significantly the measured mechanical properties. Indeed the industrial-scale synthesis process of commercial excipients, and their post-synthesis handling, differ significantly from the lab-scale synthesis of compounds A, B and C, it is thus reasonable to expect for them a more pronounced impact of re-crystallization. Finally, it must be noted that the nano-indentation technique alone is not able to give any prediction about the possible alterations of the solid-state properties of the milled material as a consequence of the received mechanical stress. This is a very important aspect to consider while designing a milling process and, to date, it is possible to evaluate it only on the milled samples, i.e. on a trial-and-error basis. For all the above-mentioned reasons, we believe nano-indentation remains of great value only in the very early development phases, when new compounds are synthesized in the order of few grams. With such limited abundance, milling process design cannot be afforded on a trial-end-error basis. However, when the synthesis scale is of many tens of grams, few milling trials, guided by the data collected in the past on other compounds, should allow to set up a decent milling process in a shorter time compared to the duration of an indentation analysis.

Despite all the limitations of nano-indentation, a correlation exists between the brittleness index $b$ and the powder behavior during milling, and it can certainly be exploited to build predictive models. The dependence of the $dv90$ or the grindability index upon $b$ deserve further investigation enriching the statistics with new compounds, especially those with little brittle fracture attitude. Another aspect to better clarify is how strongly the $dv90$ and grindability index still depend on the particle size of the feed material. The classification mechanism of jet-mills should eliminate by design the dependence of the milled product PSD on the starting material one, however this is true only in an ideal working regime which might not be met here for small $b$ compounds. With a much larger set of milling experiments it is certainly possible to define a generalized version of eq.(8) and (9), predicting the $dv90$ (as well as other characteristic diameters) as a function of the milling pressure, the feed rate and the index $b$.

We confirmed the known inadequacy of the ductile-brittle transition models by Hagen and Kendall in predicting the smallest fragment size. We stress that deeper quantitative considerations on this point cannot be made on the basis of the presented data. Working with sub-micron size particles requires, in fact, the use of different particle size characterization methods (wet dispersion, Mie scattering) and the definition of a different set of milling trials, aimed at finding the limit distribution of the different compounds rising the pressure above 6 barg when necessary.

If these points will be addressed and further investigated, we believe it will be possible to build a predictive model enabling the estimation of the PSD of a milled product solely based on the mechanical properties measured via nano-indentation. Such model remains of course valid only for the specific milling equipment used in the investigation, scaling the milling process up on larger mills or porting the process from one mill to another would require the construction of predictive models for each of them.

## Conflicts of Interest

The authors declare no conflict of interest.